\documentclass[10pt,letterpaper,twocolumn]{article}

\usepackage{usenix-2020-09}

\usepackage{hyphsubst}
\usepackage{csquotes}
\usepackage{doi}
\usepackage{listings}
\usepackage[%
capitalise,  
nameinlink,  
noabbrev,    
] {cleveref}
\usepackage{siunitx}
\usepackage{booktabs}
\usepackage[normalem]{ulem}
\usepackage{balance}
\usepackage{authblk}

\usepackage{tikz}
\usepackage{tikzpagenodes}
\usepackage{pgfplots}
\usepackage{caption}
\usepackage{subcaption}
\usepackage{paralist}
\usepackage{fontawesome}
\usepackage{tikzsymbols}
\usepackage{siunitx}

\usetikzlibrary{
  arrows,
  arrows.meta,
  backgrounds,
  bending,
  calc,
  chains,
  decorations,
  decorations.pathreplacing,
  fit,
  intersections,
  matrix,
  patterns,
  patterns.meta,
  positioning,
  scopes,
  shadows.blur,
  shapes.geometric,
  shapes.multipart
}
\usepgfplotslibrary{units, groupplots, statistics}
\pgfplotsset{compat=newest}
\colorlet{tss color}{teal!70!black}
\colorlet{trampoline color}{brown!70!black}
\colorlet{kernel color}{violet}
\lstdefinelanguage{asm}{
	morekeywords=[2]{xorl, xorq, incl, monitor, mwait, movq, mulq, imulq, addq, sysretq, wrmsr, cmpq, je, jae, jmpq},
	otherkeywords={\%rax, \%rdi, \%rsi, \%rdx, \%ecx, \%edx, \$, :, \,},
	morekeywords={\%rax, \%rdi, \%rsi, \%rdx, \%ecx, \%edx, \$, :, \,},
	morecomment=[s][\bfseries\color{gray!50!black}]{<}{>},
	morecomment=[s]{/*}{*/},
	morecomment=[l]{//},
}
\usepackage[square,comma,numbers,sort&compress]{natbib}
\newcommand{\mycomment}[3]{%
    \ifthenelse{\boolean{enablecomments}}{%
        \fbox{\bfseries\sffamily\scriptsize#1}%
        {\small\textsf{\emph{\color{#3}{\ #2}}}}}%
        {}}

\iffalse
\newcommand{\mt}[1]{\mycomment{Till}{#1}{blue}}
\newcommand{\rv}[1]{\mycomment{Viktor}{#1}{green!50!black}}
\newcommand{\MP}[1]{\mycomment{Maksym}{#1}{red!50!black}}
\else
\newcommand{\mt}[1]{}
\newcommand{\rv}[1]{}
\newcommand{\MP}[1]{}
\fi

\begin{document}

\title{New Mechanism for Fast System Calls}

\makeatletter
\renewcommand\AB@affilsepx{\protect\Affilfont}
\makeatother

\author[1]{Till Miemietz$^*$}
\author[2]{Maksym Planeta$^*$}
\author[1,2]{Viktor Laurin Reusch$^*$}
\affil[1]{
  Barkhausen Institut, Germany
}
\affil[2]{
  TU Dresden, Germany
}

\date{}

\maketitle

\begin{abstract}
  System calls have no place on the fast path of microsecond-scale systems.
  However, kernel bypass prevents the OS from controlling and
  supervising access to the hardware.
  In this paper we introduce the \emph{fastcall space}, a new layer in the
  traditional OS architecture, that hosts \emph{fastcalls}.
  A fastcall implements the fast path of a traditional kernel operation and can
  stay on the fast path, because the transition to the fastcall space is
  $\approx\times 15$ faster than to the kernel space.
  \mt{Can we still claim this after being faced with the results on AMD?}
  \rv{The meaning of \enquote{and can stay on the fast path} is not clear to
  me.}
  This way the OS does not give up the control over device access, whereas the
  applications maintain their performance.
\end{abstract}


\def\thefootnote{*}\footnotetext{These authors contributed equally to this work}\def\thefootnote{\arabic{footnote}}

\section{Introduction}

The system call layer creates a tight barrier in the transition between the user
and the kernel modes, shielding the kernel from the user for security and safety
reasons.
In a traditional system, user applications must cross this barrier to write a
file to a disk or send a network packet.
Recently however, for high-performance applications, the cost of a system call
executed on a performance critical path becomes prohibitively 
expensive~\cite{arrakis, ix, bpf_storage}.

The \emph{kernel-bypass} architectures remove the transition barrier and the
corresponding system call overhead by mapping the devices directly into the
user-space applications~\cite{dpdk, arrakis, spdk, ix, ibta, mtcp}.
Kernel bypass takes the devices away from the control of the OS and relies on the
devices to implement features like QoS~\cite{freeflow}, connection
tracking~\cite{masq}, or live migration~\cite{migros}.
Instead, we introduce the \emph{fastcall space}, a new layer within the existing
OS architectures, enabling low-overhead OS-controlled device access by user
applications.

Accessing devices through system call layer is slow for two reasons.
First, the switch between kernel and user modes has considerable performance
penalty, aggravated by recent speculation-based side-channel-attack
mitigations~\cite{spectre, meltdown}.
By itself, this aspect adds around $\SI{300}{\ns}$ (see~\Cref{sec:evaluation}) to the
\emph{datacenter tax}~\cite{kanevProfilingWarehousescaleComputer2015, arrakis}
for each system call invocation.
Second, the hot path in the Linux kernel~\cite{linux} can be long and have
several latency-inflating asynchronous calls requiring multiple intra-kernel
context switches.
For example, a write to a file goes through file system and block layers, before
landing with the device controller driver.
If the application intends to use a modern PCIe~4~$\times{}16$ device close to
its maximum request processing rate, it requires an alternative to the
system-call path.\rv{Do we have a reference for this maybe?}

Therefore, direct device access is necessary for a high performance application
but has multiple prerequisites.
First, the device becomes responsible for sharing the resources between multiple
user applications like, for example, in RDMA-networks~\cite{ibta}.
Second, if such device support is not available, kernel bypass can be used only
when the application is trusted to have full device access.
The first model requires additional device capabilities, increasing the cost of
the device and increasing the time span for deploying new
features~\cite{NetworkingToeWiki, accelnet}.
The second model is viable only for a subset of infrastructure-level software,
limiting the applicability of kernel bypass in practice.
Fastcalls combine the features of system calls and kernel bypass to offer
cost-efficient access to the hardware without limiting the applicability of
kernel bypass.
%

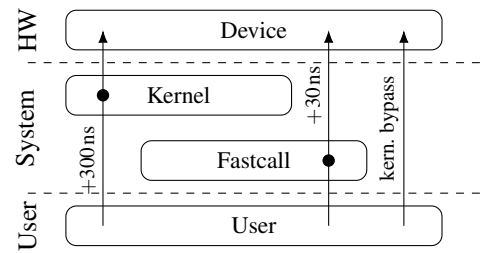
\begin{figure}
  \centering
  \begin{tikzpicture}[
  block/.style = {
    draw,
    rounded corners,
    node distance=2em,
    minimum height=1.5em,
    minimum width=5cm,
    inner sep=0,
    font=\small,
  }
]
\node[block] (device) {Device};
\coordinate[node distance=1.7em, below = of device] (c);
\node[block, minimum width=3cm, anchor=west, xshift=0cm] (kernel) at (c-|device.west) {Kernel};
\coordinate[node distance=1.7em, below = of kernel] (c);
\node[block, anchor=west, minimum width=3cm, xshift=1cm] (fastcall) at (c-|device.west) {Fastcall};
\coordinate[node distance=1.7em, below = of fastcall] (c);
\node[block, anchor=west] (user) at (c-|device.west) {User};

\path[draw, xshift=0.5cm, -{Latex}] ($(user.west)+(0.5cm,0)$) 
  -- ($(kernel.west)+(0.5cm,0)$) 
    node[midway, rotate=90, yshift=0.2cm] {\footnotesize $+\SI{300}{\ns}$} 
    node[draw, inner sep=0, circle, minimum size=0.15cm,fill] {}
  -- ($(device.west)+(0.5cm,0)$) ;
\path[draw, xshift=0.5cm, -{Latex}] ($(user.west)+(3.5cm,0)$) 
  -- ($(fastcall.west)+(2.5cm,0)$) node[draw, inner sep=0, circle, minimum size=0.15cm,fill] {}
  -- ($(device.west)+(3.5cm,0)$) 
    node[midway, rotate=90, yshift=0.2cm] {\footnotesize $+\SI{30}{\ns}$} ;
\path[draw, xshift=0.5cm, -{Latex}] ($(user.west)+(4.5cm,0)$) 
  -- ($(device.west)+(4.5cm,0)$) 
    node[midway, rotate=90, yshift=0.2cm] {\footnotesize kern. bypass};

\coordinate (c1) at ($(fastcall)!0.5!(user)$);
\coordinate (c2) at (c1-|user.west);
\coordinate (c3) at (c1-|user.east);
\coordinate (ukb) at ($(c2) - (0.5cm,0)$);
\path[draw, dashed] (ukb) --  ($(c3) + (0.5cm,0)$);
\coordinate (c1) at ($(device)!0.5!(kernel)$);
\coordinate (c2) at (c1-|device.west);
\coordinate (c3) at (c1-|device.east);
\coordinate (kdb) at ($(c2) - (0.5cm,0)$);
\path[draw, dashed] (kdb) --  ($(c3) + (0.5cm,0)$);

\node[rotate=90] at ($(ukb)!0.5!(kdb)$) {System};
\node[rotate=90] at (ukb|-user) {User};
\node[rotate=90] at (kdb|-device) {HW};
\end{tikzpicture}
  \caption[Fastcall system layer]{Fastcall system layer. Accessing a device
    through a system call adds at least $\SI{300}{\ns}$ to the end-to-end
    latency, comparing to kernel bypass. A fascall adds
    $\approx\SI{30}{\ns}$.
  }
  \label{fig:fastcall-intro}
\end{figure}

The fastcall layer resides in a privileged domain but avoids fully entering
the kernel space (see~\cref{fig:fastcall-intro}).
Therefore, invoking a fastcall is much faster than invoking a system call.
This speedup that fastcalls provide comes at the cost of some 
limitations though, like not being able to access sensitive information inside
the kernel~\cite{spectre, meltdown}.
The fastcall layer hosts multiple \emph{fastcall functions}, code snippets
tailored to specific use cases, that represent fast paths of privileged
operations.
Fastcall functions are provided and trusted by the OS, and hence they are 
allowed to execute in the privileged CPU mode and to have direct access to 
hardware.
%
%

One way to look at fastcalls is \emph{CPU-onloading}, i.e. moving work from the
hardware accelerators to the CPU with the purpose of reducing data movement.
Unnecessary data movement has been often identified as a source of performance
bottlenecks~\cite{kanevProfilingWarehousescaleComputer2015,
boroumandGoogleWorkloadsConsumer2018, ivanovDataMovementAll}, therefore doing
work on a CPU may actually improve performance.
Moreover, our intention is to use fastcalls for running control logic, which is
traditionally executed most efficiently by general-purpose CPUs.
This way fastcalls can complement CPU-offloading and become part of future
multitenant environments.



\section{Background}
\label{sec:background}

Fastcalls build upon and extend the existing Linux kernel
\emph{system call} mechanism.
We implement fastcalls for the \texttt{x86-64} architecture, and correspondingly
provide details for this architecture, although fastcalls can be generalized
to other CPU and OS architectures as well.

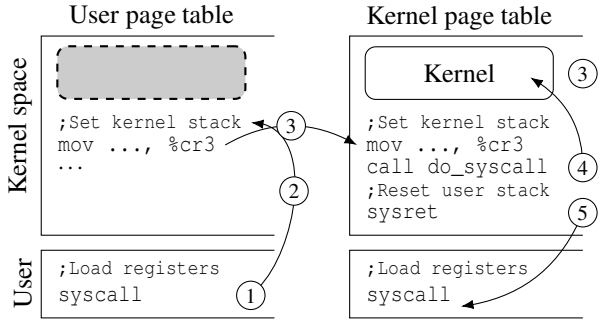
\begin{figure}
  \centering
  \begin{tikzpicture}[
  block/.style = {
    draw,
    rounded corners,
    node distance=2em,
    minimum height=1.5em,
    minimum width=5cm,
    inner sep=0,
    font=\small,
  },
  kernel active/.style = {
    draw,
    anchor=south west,
    minimum width=2.5cm,
    rounded corners, 
    minimum height=0.7cm
  },
  kernel inactive/.style = {
    kernel active,
    dashed, 
    fill=black!20,
    thick
  },
  node step/.style = {
    draw,
    circle,
    fill=white,
    inner sep=0.05cm,
    font=\footnotesize,
  }
]

\begin{scope}
\path[draw] (3.1cm, 1.1cm) -- (0, 1.1cm) -- (0, 3.75cm) -- (3.1cm, 3.75cm);
\path[draw] (3.1cm, 0) -- (0, 0) -- (0, 0.9cm)-- (3.1cm, 0.9cm);

\node[anchor=west] at (0.1cm, 0.30cm) (syscall user in) {\small \texttt{syscall}};
\node[anchor=west] at (0.1cm, 0.65cm) {\footnotesize \texttt{;Load registers}};
\node[anchor=west] at (0.1cm, 2.60cm) (kern enter) {\footnotesize \texttt{;Set kernel stack}};
\node[anchor=west] at (0.1cm, 2.30cm) (cr3 out) {\small \texttt{mov \dots, \%cr3}};
\node[anchor=west] at (0.1cm, 2.00cm) {\small \dots};
\node[kernel inactive] (kern) at (0.2cm, 2.9cm) {};
\node[node distance=.1cm, above=of kern] {User page table};
\node[rotate=90, anchor=south west, minimum width=2.9cm] at(0, 1.1cm) {Kernel space};
\node[rotate=90, anchor=south west, minimum width=0.9cm] at(0, 0cm) {User};

\draw[-{Latex}] ($(syscall user in.east-|kern enter.east)$) 
    to[rounded corners] node[node step, near start] {1} (syscall user in-|kern enter.east)
   to[rounded corners, in=0] node[node step] {2} ($(kern enter.east)+(0.0cm, 0)$);
\end{scope}

\begin{scope}[xshift = 4.1cm]
\path[draw] (3.1cm, 1.1cm) -- (0, 1.1cm) -- (0, 3.75cm) -- (3.1cm, 3.75cm);
\path[draw] (3.1cm, 0) -- (0, 0) -- (0, 0.9cm)-- (3.1cm, 0.9cm);

\node[anchor=west] at (0.1cm, 0.30cm) (syscall user in) {\small \texttt{syscall}};
\node[anchor=west] at (0.1cm, 0.65cm) {\footnotesize \texttt{;Load registers}};
\node[anchor=west] at (0.1cm, 2.60cm) (kern enter) {\footnotesize \texttt{;Set kernel stack}};
\node[anchor=west] at (0.1cm, 2.30cm) (cr3 in) {\small \texttt{mov \dots, \%cr3}};
\node[anchor=west] at (0.1cm, 2.00cm) (do syscall) {\small \texttt{call do\_syscall}};
\node[anchor=west] at (0.1cm, 1.70cm) (kern enter) {\footnotesize \texttt{;Reset user stack}};
\node[anchor=west] at (0.1cm, 1.40cm) (sysret) {\small \texttt{sysret}};

\node[kernel active] (kern) at (0.2cm, 2.9cm) {Kernel};
\node[node distance=.1cm, above=of kern] {Kernel page table};

\draw[-{Latex}] (cr3 out.east) to[bend left] node[node step] {3} (cr3 in.west);
\node[node step] (lbl) at (3.1cm, 3.25cm) {3};
\draw[-{Latex}]  (do syscall.east-|lbl)  to [bend right]  (2.4cm, 3.2cm);
\node[node step] at (do syscall.east-|lbl) {4};
\draw[-{Latex}]  (sysret.east-|lbl)  to [bend left]  ($(syscall user in.south east)+(0,0.1cm)$);
\node[node step] at (sysret.east-|lbl) {5};
\end{scope}

\end{tikzpicture}
  \caption[Address spaces]{An address space consists of a kernel and a user space.
    As part of KPTI, the kernel maintains two sets of address spaces for each
    process and must switch between corresponding page tables, when crossing the
    privilege boundary.
    \rv{Why is there another 3 in the figure? Also, the 5 actually happens in
    the user page tables.} 
    \MP{Double 3 is intended. Need to move sysret to the right side of the
    picture.}
  }
  \label{fig:syscall-bg}
\end{figure}

\Cref{fig:syscall-bg} illustrates the specifics of the system call implementation of
the Linux kernel~\cite{entry64}.
When an application wants to perform a system call (e.\ g.\ \texttt{open},
\texttt{read}, or \texttt{write}), it moves the system call number into register
\texttt{rax} and the system call parameters into other registers.
Then, the application executes the \texttt{syscall} instruction
(\cref{fig:syscall-bg}, 1), resulting in the following operations: transfer of
the execution flow to a preconfigured kernel entry point (\cref{fig:syscall-bg},
2), transition of the CPU into privileged mode, deactivation of interrupts,
and storing the return address in the \texttt{rcx} register~\cite[pp.\ 2984,
1854]{intel_manual}.
Typical kernel operations can only be executed in the privileged mode, and the
\texttt{syscall} instruction  guarantees that there only exists a single entry
point into the kernel code, which can be firmly secured against attacks.

Immediately after entering the kernel most of the kernel memory is not valid,
because it is not mapped into the current address space
for security reasons~\cite{linux_doc_mm}.
The kernel starts by setting up the kernel stack and then changes page tables
by storing a new value into the \texttt{cr3} register (\cref{fig:syscall-bg}, 3,
\cite[p.\ 2857]{intel_manual}).
Changing page tables is a part of KPTI~\cite{kpti}, a technique for protecting
the kernel data from speculation-based side-channel attacks~\cite{meltdown} and
is not required for all CPUs.
Only after changing the page tables the rest of kernel code and data becomes
accessible, and the control goes to the actual system call handlers
(\cref{fig:syscall-bg}, 4).
\mt{I would reorder the paragraph: State earlier that KPTI is not needed for
all CPUs, and then explain why we need it and how it works.}

The system call handler identifies the exact system call by the number in the
register \texttt{rax} and then dispatches it for processing to the corresponding
kernel subsystem.
After the system call is complete, the kernel returns from the
\texttt{do\_syscall} function and performs the previously taken actions in
reverse (\cref{fig:syscall-bg}, 5): restores the original user-view of the
address space, including the stack and registers, and sets up the correct return
address.
Finally, the \texttt{sysret} instruction returns the execution flow from the
privileged CPU mode back to the user application.

It takes overall around \SI{300}{\ns} to reach the required kernel subsystem and
return from there~(see~\cref{sec:evaluation}).
This administrative overhead is prohibitive for microsecond-scale applications, especially if
the actual functional part of the system call is short~\cite{demikernel, dpdk,
arrakis}.
Unfortunately, this overhead cannot be entirely avoided, because system
calls are supposed to execute privileged operations, which normally cannot be
delegated to user mode.

The largest source of overhead are the page table switch and other
mitigations of speculation-based attacks.
We analyze the sources of the overhead in more detail in \cref{sec:evaluation}.
Further sources of overhead come from complicated kernel logic, which may not
always be optimized for the fast path.
In the next section, we describe how fastcalls avoid most of the overhead
connected to system call invocation without sacrificing security.

\mt{tbh, in the light of the results that we obtained from AMD, I am not sure
whether PTI should take so much space here...}





\section{Design}%
\label{sec:design}

Fastcalls are intended to be used similarly to system calls.
Unlike a system call, a fastcall has a very specific purpose (sending a packet a
specific destination or writing into a specific set of blocks) and covers only
the fast path.

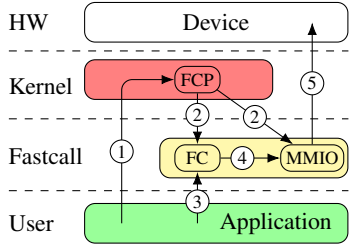
\begin{figure}
  \centering
  \colorlet{fastcall color}{yellow!40!white}
\colorlet{kernel color}{red!50!white}
\colorlet{app color}{green!40!white}

\begin{tikzpicture}[
  block/.style = {
    draw,
    rounded corners,
    node distance=2em,
    minimum height=1.5em,
    minimum width=3.5cm,
    inner sep=2pt,
    font=\small,
  },
  small block/.style = {
    block,
    inner sep=2pt,
    minimum width=0.6cm,
    minimum height=1em, 
    anchor=west,
    font=\scriptsize
  },
  space lbl/.style = {
    text width=1cm,
    align=left,
    font=\small,
    xshift=0.5cm
  },
  lbl/.style = {
    draw, midway, inner sep=1pt, circle, minimum size=0.15cm, fill=white,
    font=\scriptsize,
    midway
  },
  path/.style = {
    draw,
    rounded corners,
    -{Latex},
  },
]
\node[block] (device) {Device};
\coordinate[node distance=1.4em, below = of device] (c);
\node[block, minimum width=2.5cm, anchor=west, xshift=0cm, fill=kernel color] (kernel) at (c-|device.west) {};
\node[small block, xshift=1.2cm] (fcp) at (c-|device.west) {FCP};
\coordinate[node distance=1.7em, below = of kernel, yshift=-0.5em] (c);
\node[block, anchor=west, minimum width=2.5cm, xshift=1cm, fill=fastcall color] (fastcall) at (c-|device.west) {};
\node[small block, xshift=1.2cm] (fc) at (c-|device.west) {FC};
\node[small block, xshift=1.1cm] (mmio) at (fc) {MMIO};
\coordinate[node distance=1.7em, below = of fastcall] (c);
\node[block, anchor=west, fill=app color] (user) at (c-|device.west) {\qquad\qquad\quad Application};


\coordinate (c1) at ($(fastcall)!0.5!(user)$);
\coordinate (c2) at (c1-|user.west);
\coordinate (c3) at (c1-|user.east);
\coordinate (ukb) at ($(c2) - (1cm,0)$);
\path[draw, dashed] (ukb) --  (c3);
\coordinate (c1) at ($(device)!0.5!(kernel)$);
\coordinate (c2) at (c1-|device.west);
\coordinate (c3) at (c1-|device.east);
\coordinate (kdb) at ($(c2) - (1cm,0)$);
\path[draw, dashed] (kdb) --  (c3);
\coordinate (c1) at ($(fastcall)!0.5!(kernel)$);
\coordinate (c2) at (c1-|kernel.west);
\coordinate (c3) at (c1-|device.east);
\coordinate (fdb) at ($(c2) - (1cm,0)$);
\path[draw, dashed] (fdb) --  (c3);

\node[space lbl] at ($(ukb)!0.25!(kdb)$)  {Fastcall};
\node[space lbl] at ($(ukb)!0.75!(kdb)$) {Kernel};
\node[space lbl] at (ukb|-user) {User};
\node[space lbl] at (kdb|-device) {HW};

\path[draw, rounded corners, xshift=0.5cm, -{Latex}] ($(user.west)+(0.5cm,0)$) 
  -- ($(kernel.west)+(0.5cm,0)$) 
      node[lbl] {1}
  -- (fcp);
\path[path] (fcp) -- (fc)
  node[lbl, yshift=0.15em] {2};
\path[path] ($(fcp.south east)+(-0.05cm, 0.05cm)$) -- (mmio) node[lbl] {2};
\path[path] (user-|fc) -- (fc) node[lbl, yshift=-0.15em] {3};
\path[path] (fc) -- (mmio)
  node[lbl, xshift=-.3em] {4};
\path[path] (mmio) -- (mmio|-device) node[lbl] {5};

\end{tikzpicture}
  \caption[Fastcall invocation]{Fastcall invocation.
    %
    First, the application requests (1) the fastcall provider (FCP) to register
    a fastcall (FC).
    The FCP installs (2) the fastcall, including the MMIO mapping, into the
    application's fastcall space.
    Later, the application accesses the device by invoking (3) the fastcall, which
    accesses (4) the  MMIO region of the corresponding device and triggers (5) the
    corresponding IO operation.
    %
    Fastcalls and the MMIO region are protected against modifications by the application.
    %
    The kernel is additionally protected against side-channel attacks.
  }
  \label{fig:fcl-design}
\end{figure}

\Cref{fig:fcl-design} depicts a new layer in the OS architecture: the 
  \emph{fastcall space}.
Like the kernel, the fastcall space resides in a privileged domain, that the
user application normally has no access to.
When invoking a fastcall, similarly to a system call, the application
transitions to a well-defined entry point inside the fastcall space.
Such deterministic transition guarantees that the privileged operation is
executed in the way expected by the OS.

Unlike a system call, the fastcall entry point is intended to be as small as
possible to minimise the overhead introduced by privilege transitions.
Pursuing this goal, we even remove instructions, mitigating CPU-targeting side
channel attacks.
As a result, the fastcall space is less protected against unauthorised read
accesses resulting from application initiated side-channel attacks.
Therefore, each fastcall must be designed as if a user-application had
read access to the fastcall space.
For this reason, fastcalls cannot access the kernel state, use the kernel
stacks, or call into kernel functions.

The application starts without having access to any fastcalls.
All privileged operations must go through the kernel.
If a certain operation (e.g.\ sending a network packet) resides on a critical
path, the application may request a fastcall from a \emph{fastcall provider}.
The provider creates a fastcall and maps it into the fastcall space of the
application, if such an action matches the established security policy.

A system can have multiple fastcall providers residing in the kernel or running
as trusted user-level processes.
For our evaluation we implemented fastcall providers as kernel modules.
Only after the fastcall has been mapped into the application's fastcall space,
can it be called with minimum latency and without going into the kernel.

All fastcalls can be tuned individually to ensure a specific
security or resource usage policy at the moment of fastcall invocation.
As an example of a security policy, the fastcall may check that the
application-generated network packet header contains only a permitted 
destination IP address.
A resource usage policy may rate-limit the frequency of fastcall invocations,
therefore limiting the frequency of device accesses. 

The kernel must ensure that the fastcall is self-contained in terms of memory
accesses and executed code.
This means that the fastcall may neither perform memory operations outside a 
dedicated region nor is it allowed to execute any code that is not explicitly
part of it, like external functions.
There are two reasons for these limitations:
First, to enable the highest performance and the simplest implementation, the
fastcall may not generate CPU exceptions like page faults.
Second, all possible execution paths of a fastcall must be observable by the
fastcall provider to enable verification of the fastcall logic.
Although we do not cover fastcall verification in this paper, we expect
fastcalls to be automatically verified similarly to eBPF~\cite{ebpf} functions.

To guarantee that a fastcall causes no page faults, each fastcall receives a
block of \emph{scratchpad} memory to be used as a stack.
Additionally, the kernel creates a pinned memory region shared between the
fastcall and the application.
A part of the shared region is writable by the application, another part is only
readable.
The application can pass data to a fastcall only through CPU registers or via
the aforementioned shared memory region.

\mt{Maybe for clarity, we should further differentiate between a fastcall
(generic function) and a fastcall instance (adapted function that is mapped
into the fc-space of a user process)}


\section{Implementation} \label{sec:implementation}

We implemented the concept of fastcalls in the Linux kernel~\cite{linux}%
\footnote{The source is public: \url{https://github.com/vilaureu/linux/tree/fastcall}.
\rv{We need to think about how to publish the code.}}.
This implementation augments the system call handling of the kernel to allow the execution of low-latency fastcall functions.
The execution of fastcall functions in a higher CPU privilege mode isolates
fastcalls from applications and enables these functions to access system pages,
which user mode cannot manipulate.
Overall, this implementation of the fastcall design largely focuses on extending
kernel facilities and reducing the latency of application--kernel transitions.

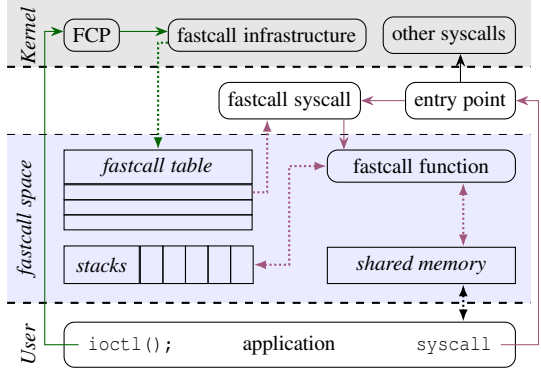
\begin{figure}
	\centering
	\colorlet{call color}{magenta!60!black}
\colorlet{reg color}{green!40!black}
\colorlet{fc space color}{blue!8!white}
\colorlet{kernel space color}{lightgray!40!white}

\begin{tikzpicture}[
	font=\footnotesize,
	caption/.style={node distance=0mm},
	inner/.style={node distance=2mm},
	block/.style={draw, align=center, node distance=0mm},
	exec/.style={block, rounded corners},
	stack/.style={block, minimum width=3mm, minimum height=5mm},
	entry/.style={block, minimum width=2.5cm, minimum height=2mm, inner sep=1mm},
	right block/.style={block, minimum width=2.5cm},
	left label/.style={anchor=south, rotate=90, font={\footnotesize\itshape}},
	lower label/.style={left label, yshift=2.5mm},
	upper label/.style={lower label, yshift=5mm},
	con/.style={-{Stealth}},
	data/.style={-{Latex[scale=0.7]}, densely dotted, thick},
	bidata/.style={data, {Latex[scale=0.7]}-{Latex[scale=0.7]}},
]

\node[exec, minimum width=6cm, minimum height=6mm] (app) {application};
\node [inner, left=of app.east] (call fastcall) {\texttt{syscall}};
\node [inner, right=of app.west] (ioctl) {\texttt{ioctl();}};

\node[stack, minimum width=1cm, above right=of app.north west, yshift=5mm] (stack0) {\textit{stacks}};
\foreach \x in {1, ..., 5} \node[stack, right=of stack0, xshift=(\x - 1) * 3 mm] (stack\x) {};

\foreach \x in {0, ..., 2} \node[entry, above right=of stack0.north west, yshift=\x * 2 mm + 2mm] (table\x) {};
\node[entry, above=of table2] (table3) {\textit{fastcall table}};

\coordinate (shared root) at (app.east |- stack0.south);
\node[right block, above left=of shared root] (shared) {\textit{shared memory}};

\coordinate (function root) at (app.east |- table3.north);
\node[right block, exec, below left=of function root] (function) {fastcall function};

\node[exec, above left=of function.north east, yshift=4mm] (entry) {entry point};
\node[exec] (syscall) at (app |- entry) {fastcall syscall};

\node[exec, above=of syscall, yshift=4mm, xshift=-3mm] (provider) {fastcall infrastructure};
\coordinate (module root) at (app.west |- provider);
\node[exec, right=of module root] (module) {FCP};
\coordinate (normal syscall root) at (app.east |- provider);
\node[exec, left=of normal syscall root] (normal syscall) {other syscalls};

\coordinate (user fc left) at ($([xshift=-1.5cm]app.north west)!0.5!(stack0.south west)$);
\coordinate (user fc right) at ([xshift=2mm]app.east |- user fc left);
\draw[dashed, thick] (user fc left) -- (user fc right);

\coordinate (user kernel) at ($(table3.north)!0.5!(syscall.south)$);
\coordinate (user kernel left) at (user fc left |- user kernel);
\draw[dashed] (user kernel left) -- (user fc right |- user kernel);

\coordinate (kernel kpti) at ($(syscall.north)!0.5!(provider.south)$);
\coordinate (kernel kpti left) at (user fc left |- kernel kpti);
\draw[dashed, thick] (kernel kpti left) -- (user fc right |- kernel kpti);

\coordinate (syscall left) at (module.west |- syscall.south);
\node[lower label] at (app.west) {User};

\node[lower label] at (module.west) {Kernel};
\node[lower label] at ($(stack0.south west)!0.5!(table3.north west)$) {fastcall space};

\begin{scope}[on background layer]
	\fill[fc space color] (user kernel left) rectangle (user fc right);
\end{scope}

\coordinate (top right) at ([yshift=2.5mm]user fc right |- provider.north);
\begin{scope}[on background layer]
	\fill[kernel space color] (kernel kpti left) rectangle (top right);
\end{scope}

\draw[con, call color] (call fastcall.east) -- ++(0.5, 0) |- (entry.east);
\draw[con, call color] (entry) -- (syscall);
\coordinate (syscall function) at ($(syscall.south east)!0.5!(function.north west)$);
\draw[con, call color] (syscall.south -| syscall function) -- (function.north -| syscall function);
\draw[data, call color] (table2) -| ([xshift=-3mm]syscall.south);
\coordinate (function shared) at ([xshift=-7mm]function.south east);
\draw[bidata, call color] (function shared) -- (shared.north -| function shared);
\draw[bidata, call color] (function.west) -- ++(-0.5, 0) |- (stack5.east);

\draw[con, reg color] (ioctl.west) -- ++(-0.45, 0) |- ([yshift=0.5mm]module.west);
\coordinate (module provider) at ([yshift=0.5mm]module);
\draw[con, reg color] (module.east |- module provider) -- (provider.west |- module provider);
\draw[data, reg color] ([yshift=-1mm]provider.west) -| (table3.north);

\draw[bidata] (shared.south -| function shared) -- (app.north -| function shared);
\draw[con] (entry) -- (entry |- normal syscall.south);
\end{tikzpicture}
	\caption[Control and data flow in the fastcall implementation]{
		\underline{Control} and \dotuline{data flow} of {\color{reg color} registrations} and {\color{call color} invocations} in the fastcall implementation.
		{\color{reg color} Registrations} via \texttt{ioctl} go through a
		fastcall-provider (FCP) module and the built-in fastcall infrastructure,
		where they are mapped and inserted into the fastcall table.
		Each \texttt{syscall} instruction {\color{call color} invocation} will be
		forwarded either to \colorbox{kernel space color}{kernel} or
		\colorbox{fc space color}{fastcall} space.} \label{fig:implementation}
\end{figure}

The high-level view of the fastcall implementation is depicted in \cref{fig:implementation}.
The flow of managing fastcalls is shown on the left.
Authorized applications register for fastcalls via fastcall providers, which are loadable kernel modules.
They then interact with the built-in fastcall infrastructure in the kernel.
Modules allow administrators to extend systems with new functionality and aid the rapid development of new fastcalls.

The right side of \cref{fig:implementation} depicts the invocation of fastcalls through the normal system call interface.
\Cref{fig:entry-listing} provides a more detailed view of this invocation.
To minimize fastcall latency, fastcalls are detected right at kernel entry and
forwarded to the \emph{fastcall dispatcher}.
The dispatcher locates the desired fastcall function in memory and executes it.
All operations in the fastcall space run in the privileged mode, meaning that
the fastcalls have kernel-level capabilities.

The fastcall space hosts all the facilities relevant for executing fastcall
functions as shown in \cref{fig:implementation}.
One of these facilities is the per-process fastcall table.
Entries in this table are created by the fastcall infrastructure and they contain pointers to fastcall functions, to data regions and configuration parameters which are passed through to fastcall functions.
The fastcall system call uses this table to locate the fastcall functions in fastcall space, which is depicted in \cref{fig:entry-listing}.
In this sense, the fastcall table acts a table of capabilities to fastcall
functions, just like the file descriptor table contains capabilities to file
descriptors.

The fastcall space hosts the stacks for the fastcall functions, which is
necessary to write fastcalls in C.
To avoid issues with concurrently running fastcalls, the stacks are assigned per
CPU and interrupts are kept disabled during the fastcall execution.
The fastcall space also hosts the shared memory regions to exchange data with
user applications and hosts device memory mappings and fastcall-private pages (e.\
g.\ for locks and counters).
In general, these facilities enable fastcalls to perform privileged operations without fully entering the kernel, which results in reduced latency compared to system calls.

\begin{figure}
	\input{fastcall-listing.tex}
	\caption[Fastcall entry and exit]{Fastcall entry and exit.
	System calls enter the kernel at label \texttt{entry\_SYSCALL\_64}.
	Lines 2 and 3 test for the special fastcall system call number and branch
	off to the fastcall dispatcher if this is the case.
	Lines 7 to 11 compute the address of a fastcall function inside the fastcall
	table using the fastcall number and the fixed address of the table.
	Line 12 jumps to the specific fastcall.
	Finally, the fastcall function returns to the application via
	\texttt{sysret} in line 17.} \label{fig:entry-listing}
\end{figure}

Fastcalls reduce overhead of system calls by not mitigating side-channel attacks
like \emph{Meltdown}~\cite{meltdown}.
Meaning that, due to KPTI, the kernel
memory is not mapped when fastcalls run.
As a result, fastcalls cannot access most of the kernel memory.
Instead, we allocate a portion of the process address space to fastcall space.
The fastcall space memory cannot be modified directly from the user mode (except
shared memory, see \cref{fig:implementation}) and the layout cannot be altered
by regular memory-management system calls.
All modifications to the fastcall space layout must instead go through the in-kernel
fastcall infrastructure.
Else, malicious applications could, for example, exchange mapped device memory
and trick fastcalls into writing to the wrong device.
Therefore, the concept of fastcall space protects fastcalls from malicious applications even though memory is mapped to user space.

Moreover, the fastcall space exhibits a special behavior on forks~\cite{fork}.
Normally, a fork would make a one-to-one copy of memory pages, breaking
potentially fastcall data structures, e.g. locks, currently in active use by
fastcalls.
Such inconsistencies should not occur in kernel mode, therefore we decided to
reset the complete fastcall space of children upon forks.


\section{Evaluation} \label{sec:evaluation}
We conduct a set of microbenchmarks to compare the latency of fastcall function
invocations to alternative approaches: \textit{vDSO} functions, system calls,
and \texttt{ioctl} handlers.
The source code of all
benchmarks\footnote{\url{https://github.com/vilaureu/fastcall-benchmarks}}\footnote{\url{https://github.com/vilaureu/linux/tree/fccmp}} is
publicly available online.
\rv{How do we want to handle source code? Maybe we should create some proper repository for our project.}

We have chosen \textit{vDSO}~\cite{vdso_man}, because it indicates the lower execution
time boundary for kernel-provided functionality.
However, vDSO does not allow kernel state modifications, so fastcalls
remain a more powerful tool.
In contrast, \texttt{ioctl} handlers offer full kernel functionality, but incur
the highest overhead among the aforementioned approaches.

All of the evaluated techniques are compared in three different scenarios: the execution of no-operation functions, the copying of small arrays, and the invocation of non-temporal copies.
In the first scenario, the called functions directly return back to the calling application.
The other two scenarios show how the latencies of the tested mechanisms compare when they actually perform some work.
In the array-copying scenario the called functions copy a small string of 64 byte into an internal buffer, which could later on be used to manipulate kernel data.
Because this scenario is very simple, the copy operations will primarily work on the CPU caches.
The non-temporal copy scenario prevents this by writing 64 bytes directly to main memory using the \texttt{vmovntdq} \cite[p.~1273]{intel_manual} vector instruction.
The concrete 64-byte value originates from the size of commands in NVMe submission queues \cite[p.~65]{nvme}.
This benchmark scenario helps to answer the question of how relevant the lower latency of the fastcall mechanism is when working with PCIe devices.
All in all, these microbenchmarks allow to evaluate how the fastcall mechanism compares to the other approaches under different load scenarios.

Finally, as a third variable, we change the mitigation settings for CPU vulnerabilities.
All benchmarks are performed once with the default set of mitigations (\texttt{mitigations=\linebreak[0]auto} \cite{cmd_params}), once without \textit{KPTI} (\texttt{nopti}), and once with all mitigations disabled entirely (\texttt{mitigations=\linebreak[0]off}).
This differentiation is important because countermeasures against \textit{Meltdown}, \textit{Spectre}~\cite{spectre}, and \emph{Microarchitectural Data Sampling (MDS)}~\cite{mds} had a noticeable effect on the system call latency when tested on our benchmark system.
Overall, depending on the enabled mitigations, the measured latencies of the functions under test might differ and are worth investigating.

\begin{table}
\caption{System Configuration}
\renewcommand{\arraystretch}{1.1}
\begin{tabular}{p{.12\textwidth} p{.3\textwidth}} \toprule
	Component & Configuration \\ \midrule
	CPU & Intel Core i7-4790 CPU at \SI{3.6}{\giga\hertz}  \\
	Memory & \SI{32}{GiB} with 4 DDR3-1600 DIMMs \\
	Kernel & based on Linux v5.11~\cite{kernel_v5.11} \\
	Distribution &  Debian Testing \textit{Bullseye}~\cite{bullseye} \\
	Settings & performance CPU governor~\cite{governor}, no~HT~\cite{hyper-threading}, no~Turbo Boost~\cite{turbo_boost} \\ \bottomrule
\end{tabular}
\label{tab:setup}
\end{table}

\cref{tab:setup} lists basic information about our benchmark system, which we
tuned for low runtime variance.
The benchmarks themselves are regular applications executing in user mode and
use the microbenchmark library \textit{benchmark} \cite{benchmark}.
This library executes the code of interest in tight loops until the average
execution times become statistically stable.
We also independently verified that latency variation for all our benchmarks was
negligible (IQR of \SI{3.3}{\nano\second} for an empty system call and 0 for a noop
fastcall).

\subsection{Results}

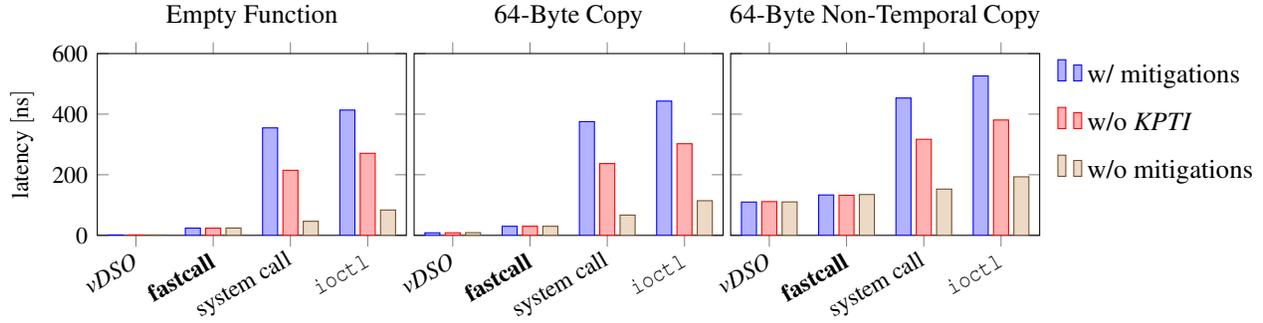
\begin{figure*}
\centering
\begin{tikzpicture}[font=\small]
\begin{groupplot}[
group style={
	group size=3 by 1,
	horizontal sep=1mm,
	vertical sep=4mm,
},
width=0.32\textwidth,
height=4cm,
xmin=-0.5,
xmax=3.5,
xtick={0, 1, 2, 3},
xticklabels={\textit{vDSO}, \textbf{fastcall}, system call, \texttt{ioctl}},
xticklabel style={
	xshift=1mm,
	yshift=-1mm,
	rotate=30,
	anchor=east,
},
ymin=0,
ymax=600,
title style={font=\normalsize},
]
	\nextgroupplot[
	title={Empty Function},
	ybar,
	ylabel={latency},
	y unit={\si{\nano\second}},
	bar width=2mm,
	]
		\addplot table[x=x, y=kpti] {benchmarks/noop.csv};
		\addplot table[x=x, y=no-kpti] {benchmarks/noop.csv};
		\addplot table[x=x, y=no-miti] {benchmarks/noop.csv};

	\nextgroupplot[
	title={64-Byte Copy},
	ybar,
	yticklabels={,,},
	bar width=2mm,
	]
		\addplot table[x=x, y=kpti] {benchmarks/array.csv};
		\addplot table[x=x, y=no-kpti] {benchmarks/array.csv};
		\addplot table[x=x, y=no-miti] {benchmarks/array.csv};

	\nextgroupplot[
	title={64-Byte Non-Temporal Copy},
	ybar,
	yticklabels={,,},
	bar width=2mm,
	legend entries={w/ mitigations, w/o \textit{KPTI}, w/o mitigations},
	legend style={
		cells={align=left},
		draw=none,
		legend cell align=left,
		legend pos=outer north east,
		row sep=2mm,
		font=\normalsize,
	},
	]
		\addplot table[x=x, y=kpti] {benchmarks/array-nt.csv};
		\addplot table[x=x, y=no-kpti] {benchmarks/array-nt.csv};
		\addplot table[x=x, y=no-miti] {benchmarks/array-nt.csv};
\end{groupplot}
\end{tikzpicture}
\caption[Latency comparison of fastcall functions with alternatives]{%
Latency comparison of fastcall functions with alternative approaches under different scenarios.
Throughput is measured on a single CPU core with invocations in tight loops.
\MP{Add labels with the specific numbers.}
}
\label{fig:benchmarks}
\end{figure*}

The results of our benchmarks are shown in \cref{fig:benchmarks}.
When comparing all four approaches in executing a no-operation function with mitigations enabled, \textit{vDSO} and fastcall functions have by far the lowest execution times.
The mere execution of the function calls to the \textit{vDSO} takes only \SI{1.4}{\nano\second}.
Performing ring transitions for executing fastcalls increases the latency to \SI{23.9}{\nano\second}.
Nevertheless, this is still approximately 15 times faster than the \SI{354.7}{\nano\second} overhead of system calls.
For \textit{ioctl} handlers this value is even larger at \SI{413.6}{\nano\second}.
This shows, that the flexibility of \texttt{ioctl} handlers comes at the cost of increased latency.
In general, the fastcall mechanism allows to avoid a large portion of the overhead of system calls and \texttt{ioctl}.

However, this aspect changes when considering the other mitigation settings.
The mitigation settings do not affect fastcalls because they skip the kernel entry sequence, which contains the relevant mitigations.
This also applies to \textit{vDSO} functions, which are by design executed in user mode.
On the other hand, the mitigation settings heavily influence the system call and \texttt{ioctl} approaches.
System calls only show \SI{46.4}{\nano\second} of latency when all mitigations are disabled.
Hence, in the second scenario they come close to the \SI{24.1}{\nano\second} of
the fastcall mechanism.
This shows the huge cost of mitigations for microsecond-scale systems.
Given that security is a necessity in a multi-tenant environment, the rest of
the evaluation focuses on the fully-mitigated systems.

The second and the third plot in \cref{fig:benchmarks} show the four approaches in more practical scenarios.
The fastcall mechanism still holds an advantage under these conditions.
For example, fastcall functions are about $\times 12.5$ faster in the array-copy
scenario and $\times 3.4$ faster in the non-temporal-copy scenario. 
This demonstrates that even relative to the latency of main memory accesses, the
overhead of the fastcall mechanism is relatively small.
This might imply that fastcall functions could even compete with kernel-bypass
techniques in regard to overall latency.

\subsection{Auxiliary Benchmarks}

\begin{figure}
\begin{tikzpicture}[
	font=\small
]
\begin{groupplot}[
	group style={
		group size=2 by 1,
		horizontal sep=2.2cm,
		ylabels at=edge left,
	},
	height=4.5cm,
	xticklabel style={
		xshift=1mm,
		yshift=-1mm,
		rotate=30,
		anchor=east,
	},
	ymin=0,
	ylabel={latency},
	y unit={\si{\micro\second}},
]
\nextgroupplot[
	ybar stacked,
	width=3cm,
	xmin=-0.5,
	xmax=1.5,
	xtick={0,1},
	xticklabels={Registration, Deregistration},
	bar width=2mm,
	legend entries={w/o\\mappings, w/\\mappings},
	legend reversed=true,
	legend style={
		cells={align=left},
		draw=none,
		legend cell align=left,
		legend pos=outer north east,
		row sep=2mm,
	},
]
	\addplot coordinates {
	        (0, 1.383) (1, 2.354)
	};
	\addplot coordinates {
	        (0, 0.793) (1, 1.835)
	};
\nextgroupplot[
	ybar stacked,
	width=2.5cm,
	xtick={0},
	xticklabels={\texttt{Fork}ing},
	bar width=2mm,
	legend entries={stock kernel, w/o\\registrations, w/\\registrations},
	legend reversed=true,
	legend style={
		cells={align=left},
		draw=none,
		legend cell align=left,
		legend pos=outer north east,
		row sep=2mm,
	},
]
	\addplot coordinates {
	        (0, 38.055)
	};
	\addplot coordinates {
	        (0, 0.674)
	};
	\addplot coordinates {
	        (0, 4.956)
	};
\end{groupplot}
\end{tikzpicture}
\caption[Plots depicting the control path overhead of the fastcall mechanism]{Bar charts depicting the median latencies of fastcall function registration and deregistration as well as for \texttt{fork}ing processes. (De)registration is tested with and without mapping auxiliary memory areas. \texttt{Fork}s are compared between the modified fastcall kernel and an unmodified v5.11 Linux kernel. Additionally, \texttt{fork}ing with 100 fastcall functions registered beforehand is evaluated.}
\label{fig:control_path}
\end{figure}
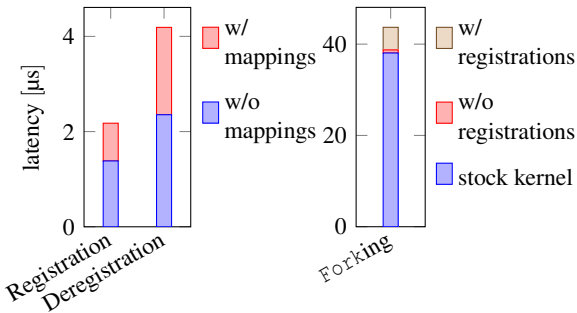

This section discusses some auxiliary benchmarks, which cover latencies of the control path of the fastcall mechanism, the general process flow, and overall system performance.
These benchmarks are performed by using the \texttt{clock\_gettime} \textit{vDSO} functionality while keeping all default mitigations enabled.
The first plot in \cref{fig:control_path} depicts the latencies involved with registering fastcall functions.
This takes $1.4$ to \SI{2.6}{\micro\second} depending on the number of additional mappings created along with the text segment.
The deregistration only takes a couple of microseconds too: $2.4$ to \SI{4.8}{\micro\second}.
Because applications are expected to register fastcall functions very infrequent, the control-path overhead of the new mechanism is likely not adverse in any normal use case.

Another aspect of interest is the influence of the fastcall mechanism on the latency of forks because this approach requires modifications to the \texttt{fork}~\cite{fork} handler of the kernel.
More precisely, the modified kernel removes fastcall memory mappings from the child processes and recreates fresh fastcall tables for them.
The latter operation even applies to processes which do not use the fastcall mechanism.
This might affect the overall system performance.
\cref{fig:control_path} only shows a slight increase in \texttt{fork} latency between the unchanged, stock kernel and the modified one.
The influence on the overall system performance should therefore be minuscule.
A more noticeable overhead only arises if the fastcall mechanism is heavily utilized.
But even this overhead might be avoidable by using the related \texttt{vfork} system call \cite{vfork}, which does not trigger a reset of the fastcall table.
In general, these benchmarks have shown that the control flow of the fastcall mechanism does not introduce much overhead into applications and does not compromise the responsiveness of the system as a whole.
\MP{Regarding red bar: How much overhead, is it statistically significant, where does it come from?}
\rv{The small overhead of forks in the fastcall kernel comes from the recreation of the fastcall table. Because I use the Google benchmark library, I only have averages and can not compute the significance. But forks on the fastcall kernel were consistently a bit slower.}


\section{Related Work}%
\label{sec:related}

Previous work improved the traditional POSIX API by accelerating
privilege transition, reducing the number of system calls, removing unnecessary
memory copies, or improving application wake-up time.
In this section we put fastcalls in the context of some of these techniques.

%
%

The most straightforward way to improve system call performance is to make the
privilege transition faster.
Currently the common way to make a system call is the \texttt{syscall} instruction,
which replaced a slower software interrupt mechanism~\cite{intel_manual}.
%
%
Call gates~\cite{intel_manual}, despite initial performance
advantages~\cite{itanium}, have been hard to use and did not gain momentum.
Moreover, call gates do not offer performance advantages
for modern CPUs anymore~\cite{lotrx86}.
Intel recently proposed FRED~\cite{fred}, an extension to the x86-64 architecture.
FRED is designed to improve the performance or ring transitions and unify
system call and interrupt invocation mechanisms.
\emph{SkyBridge} instead utilizes the \texttt{vmfunc} instruction to transition 
to another process without the overhead of standard context switches, 
but in return weakens OS security guarantees~\cite{skybridge}.
The fastcall layer also could employ any of the aforementioned methods to transition
into fastcall space, but we considered our method to be the most practical.
\MP{Looks too much detail to me. Leave for arxiv?}
\mt{imo we could drop the sentences on syscall and call gates entirely}

Modern microsecond-scale systems often choose to avoid system calls entirely.
Linux' vDSO mechanism maps system calls directly into the application,
eliminating the need for costly privilege transitions~\cite{vdso_man}.
This mechanism offers only few system calls, that can only access non-secret
read-only state of the kernel.
The fastcall mechanism extends the general concept of vDSO by introducing a way
to execute functions in privileged mode, allowing to manipulate kernel state in
a secure way.
\MP{Again too much detail, and vDSO $\leftrightarrow$ microsecond-scale? RLY? }

The \emph{io\_uring} interface~\cite{io_uring_intro} utilizes shared memory
buffers to speed up the exchange of I/O requests between applications and the
kernel.
With io\_uring, a system call is only required to update the kernel about 
changes of the queue state, similarly to other system call batching 
techniques~\cite{flexsc, anycall, scone}.
In a fully asynchronous mode, a kernel thread actively polls on shared memory for
new requests, thus eliminating the kernel entry and exit overhead completely.
However, polling on both the application and the kernel side leads to high CPU
utilization. 
With fastcalls instead, the application logic and the communication with I/O
devices are carried out in the same thread.

To avoid system calls one also could deploy user code directly into the
corresponding kernel subsystem.
Multiple methods employed \emph{eBPF}~\cite{ebpf} to offload simple request
processing routines from the user application into the
storage~\cite{bpf_storage}, network~\cite{xdp, bmc}, or scheduling~\cite{ghost}
subsystems of the Linux kernel.
Additionally, it is possible to offload the request processing
routines to the device~\cite{accelnet, spin, kopi}, reducing data movement and
CPU load.
These architecture improve request response latency, but do not address the
overhead of creating new requests.
In this regard, fastcalls are complementary to the aforementioned offloading
techniques, because they enable low-latency user-initiated interactions with the
kernel.

An application can get maximum performance by getting direct access to the
underlying device, bypassing the kernel.
In contrast to traditional OS architectures, \emph{kernel bypassing} techniques
employ special APIs to avoid privilege transitions and data copying between user
and kernel space.
For issuing device commands, user applications write to the corresponding device
registers directly.
Latency of event delivery is minimized, because user applications poll on the
device completion queues, instead of waiting in a blocking system call (e.~g.~\texttt{epoll}).
Such approaches exist for networking~\cite{ibta, netmap, dpdk} as well as for
storage devices~\cite{spdk, nvmedirect}.
However, removing the kernel from the data path often requires special support
from the device~\cite{ibta, mikejacksonPCIExpressTechnology2012} and results in
security and manageability concerns, because the OS loses the control over 
how the device is used.
We hope that the introduction of the fastcall space into the kernel-bypass
architectures may bring the OS back into the picture, without adding any
significant overhead.

The concept of kernel bypass has become a first-class citizen in
dataplane~\cite{arrakis, ix, demikernel} and microkernel-based~\cite{exokernel,
l4re} operating systems.
In addition to passing the device under the control of a user process, these
operating systems focus on convenient programming abstractions to reduce
development effort, improve portability, and enable resource sharing.
We see fastcalls as an instrument for replacing kernel bypass and for enabling
low-overhead kernel interposition with the purpose of passing the device to the
user application in controllable manner.

Being the most popular, the dataplane OS is not the only architecture trying to
revamp system calls. \rv{(Repeated) reference for \emph{dataplane OS} would be nice here.}
Singularity~\cite{singularity} and RedLeaf~\cite{redleaf} avoid CPU-enforced
address space isolation by relying on language guarantees and runtime checks.
Lee et al.\ employ call-gates to implement a new privileged user level for
processing sensitive data using x86 privilege rings~\cite{lotrx86},
whereas \citeauthor{klos} employ a new low-latency protection domain switching
mechanism to remove the kernel almost entirely~\cite{klos}.
Unfortunately, the latter mechanism comes with probabilistic security guarantees.
\mt{imo, the related section is too long and lacks a central theme that 
connects the listed approaches smoothly.}

\section{Discussion} \label{sec:discussion}

We believe that the extremely low overhead that fastcalls offer may bring
qualitative changes to the future OS architectures.
The kernel may offload fastpath operations automatically into the fastcall space
to improve common case performance.
In a way such offloading will take the opposite direction of how eBPF is used
nowadays.
In our vision, instead of having a 30-instruction receive routine of a TCP
packet~\cite{vanjacobsonFwdTCP30}, a fastcall will contain a similarly small
send routine.
The remainder of this section describes our vision of fastcalls im more detail.

\subsection{Fastcall model}%
\label{sec:disc-model}

To understand the decisions driving the fastcall design, we outline how we
reason about them.
A fastcall is designed to improve performance in comparison to system calls in
exchange for a more limited usage model.
This means that if overhead of a long-running system call is negligible,
implementing the same functionality in fastcall will not be useful.
Hence, a fastcall must not run longer than it is defined by the overhead
introduced by a system call.

To get a more specific estimation of how long a fastcall can run, assume any
overhead less than $O=5\%$ of the total operation to be negligible, system call
overhead to be $o_s=\SI{300}{\ns}$, and fastcall overhead to be
$o_f=\SI{30}{\ns}$.
The total runtime of operation is $T=w + o$, where $w$ is the useful work
performed by the operation, and $o$ is the overhead.
Assuming the system call and fastcall implementations are equivalent, $w$
component will remain the same.
Then the fastcall is beneficial only if:

\[
    \frac{o_s}{T} = \frac{o_s}{o_s + w} > O
\]

Solving for $w$, we get that the useful work must not take more than
$\SI{5.7}{\us}$.
If we additionally require the fastcall overhead to be negligible, the useful
work should take at least $\SI{570}{\ns}$.
These limits includes all the work required to complete a privileged operation,
including the preparatory work happening outside of fastcall or kernel space.
The lower limit is not a hard limit, but rather an indication that another
solution, possibly a hardware-based one, can offer tangible benefits, if the
overhead can be kept even lower.

To put these numbers in context, a high-end Infiniband NIC provides as low as
$\SI{600}{\ns}$ back-to-back latency~\cite{nvidiaConnectX6VPICard2020}.
With multiple switches in between, one can expect several microseconds latency in 
real world applications.
These numbers match well our expectations about fitting fastcall use case.
We observe similar situation with modern NVMe storage.

From the design point of view having a limit on the maximum execution time opens
up many interesting opportunities.
For example, we envision that in the future, fastcalls will be automatically 
verified to contain no bugs or known vulnerabilities before mapping them to the
application.
The maximum execution time of several microseconds may be sufficiently short to
even employ techniques like symbolic execution~\cite{klee} during the 
verification process.

\subsection{Fastcalls Are for Fast Paths}

Fastcalls have several implementation limitations that follow from our design
philosophy.
Specifically, a fastcall has the following limitations: It must not generate
exceptions, it must not access arbitrary user or kernel memory, and it must not
contain secrets. \rv{These points were already discussed in previous sections. This subsection is a bit redundant.}

Exceptions, like a page fault, would be detrimental for the fastcall performance
and render them useless.
If an exception is to be expected, one should simply make a conventional system
call, and get access to full kernel functionality.
Having this property as a design limitation, we plan to conduct a worst case
execution time analysis for fastcalls in the future.
Such analysis will simplify tail latency optimization at the OS level.

A corollary of the previous statement is that fastcalls cannot access user
memory, because such an access may generate a page fault.
We work around this limitation as fastcalls exchange data with the user over
shared regions of pinned memory.

Finally, because we omit side channel mitigations, when entering fastcall space,
fastcall space must not contain confidential information.
This limitation may be reconsidered for CPU architectures which offer
reliable protection of sensitive information.

\section{Conclusion}
\label{sec:conclusion}

In this paper, we explored the construction of a framework for lightweight
system calls, which we call \emph{fastcalls}. Fastcalls enable the operating
system to protect security-critical resources like kernel or device memory from
arbitrary access by applications, while having performance properties close
to that of normal function calls. 
In particular, the fastcall approach 
improves on the latency of standard system calls by up to $15\times$, while
keeping protection mechanisms against side-channel attacks in effect.
Even though fastcalls offer a limited execution environment that only allows
for running simple code snippets, they may be used to implement secure, 
fast, and CPU-efficient virtualization of devices as well as to give 
applications protected access to privileged CPU instructions. In the future, 
we plan to investigate the benefit of fastcalls for the virtualization of 
RDMA networks and NVMe-attached SSDs.

\mt{We actually never explicitly state how virtualization etc. can be done with
fastcalls...}

\bibliography{references,planeta}

\begin{thebibliography}{60}
\providecommand{\natexlab}[1]{#1}
\providecommand{\url}[1]{\texttt{#1}}
\expandafter\ifx\csname urlstyle\endcsname\relax
  \providecommand{\doi}[1]{doi: #1}\else
  \providecommand{\doi}{doi: \begingroup \urlstyle{rm}\Url}\fi

\bibitem[Net()]{NetworkingToeWiki}
networking:toe [{{Wiki}}].
\newblock URL \url{https://wiki.linuxfoundation.org/networking/toe}.

\bibitem[l4r()]{l4re}
{{L4Re}} -- {{The L4 Runtime Environment}}.
\newblock URL \url{https://l4re.org/}.

\bibitem[Arnautov et~al.()Arnautov, Trach, Gregor, Knauth, Martin, Priebe,
  Lind, Muthukumaran, O’Keeffe, Stillwell, Goltzsche, Eyers, Kapitza,
  Pietzuch, and Fetzer]{scone}
Sergei Arnautov, Bohdan Trach, Franz Gregor, Thomas Knauth, Andre Martin,
  Christian Priebe, Joshua Lind, Divya Muthukumaran, Dan O’Keeffe, Mark~L
  Stillwell, David Goltzsche, David Eyers, Rudiger Kapitza, Peter Pietzuch, and
  Christof Fetzer.
\newblock {{SCONE}}: Secure {{Linux Containers}} with {{Intel SGX}}.
\newblock page~17.

\bibitem[Belay et~al.()Belay, Prekas, Kozyrakis, Klimovic, Grossman, and
  Bugnion]{ix}
Adam Belay, George Prekas, Christos Kozyrakis, Ana Klimovic, Samuel Grossman,
  and Edouard Bugnion.
\newblock {{IX}}: A {{Protected Dataplane Operating System}} for {{High
  Throughput}} and {{Low Latency}}.
\newblock In \emph{11th {{USENIX Symposium}} on {{Operating Systems Design}}
  and {{Implementation}}}, {{OSDI}} '14, pages 49--65.
\newblock ISBN 978-1-931971-16-4.

\bibitem[Boroumand et~al.()Boroumand, Ghose, Kim, Ausavarungnirun, Shiu,
  Thakur, Kim, Kuusela, Knies, Ranganathan, and
  Mutlu]{boroumandGoogleWorkloadsConsumer2018}
Amirali Boroumand, Saugata Ghose, Youngsok Kim, Rachata Ausavarungnirun, Eric
  Shiu, Rahul Thakur, Daehyun Kim, Aki Kuusela, Allan Knies, Parthasarathy
  Ranganathan, and Onur Mutlu.
\newblock Google {{Workloads}} for {{Consumer Devices}}: Mitigating {{Data
  Movement Bottlenecks}}.
\newblock In \emph{Proceedings of the {{Twenty}}-{{Third International
  Conference}} on {{Architectural Support}} for {{Programming Languages}} and
  {{Operating Systems}}}, pages 316--331. {ACM}.
\newblock ISBN 978-1-4503-4911-6.
\newblock \doi{10/gnk2n3}.

\bibitem[Brouwer and Kerrisk()]{vfork}
Andries Brouwer and Michael Kerrisk.
\newblock vfork(2).
\newblock URL \url{https://man7.org/linux/man-pages/man2/vfork.2.html}.

\bibitem[Cadar et~al.()Cadar, Dunbar, and Engler]{klee}
Cristian Cadar, Daniel Dunbar, and Dawson Engler.
\newblock {{KLEE}}: Unassisted and {{Automatic Generation}} of
  {{High}}-{{Coverage Tests}} for {{Complex Systems Programs}}.
\newblock page~16.

\bibitem[Corbet()]{io_uring_intro}
Jonathan Corbet.
\newblock Ringing in a new asynchronous i/o {API}.
\newblock URL \url{https://lwn.net/Articles/776703/}.

\bibitem[{Daniel Firestone} et~al.(){Daniel Firestone}, {Andrew Putnam},
  {Sambhrama Mundkur}, {Derek Chiou}, {Alireza Dabagh}, {Mike Andrewartha},
  {Vivek Bhanu}, {Eric Chung}, {Harish Kumar Chandrappa}, {Somesh Chaturmohta},
  {Matt Humphrey}, {Jack Lavier}, {Norman Lam}, {Fengfen Liu}, {Kalin
  Ovtcharov}, {Jitu Padhye}, {Gautham Popuri}, {Shachar Raindel}, {Tejas
  Sapre}, {Mark Shaw}, {Gabriel Silva}, {Madhan Sivakumar}, {Nisheeth
  Srivastava}, {Anshuman Verma}, {Qasim Zuhair}, {Deepak Bansal}, {Doug
  Burger}, {Kushagra Vaid}, {David A. Maltz}, and {Albert Greenberg}]{accelnet}
{Daniel Firestone}, {Andrew Putnam}, {Sambhrama Mundkur}, {Derek Chiou},
  {Alireza Dabagh}, {Mike Andrewartha}, {Vivek Bhanu}, {Eric Chung}, {Harish
  Kumar Chandrappa}, {Somesh Chaturmohta}, {Matt Humphrey}, {Jack Lavier},
  {Norman Lam}, {Fengfen Liu}, {Kalin Ovtcharov}, {Jitu Padhye}, {Gautham
  Popuri}, {Shachar Raindel}, {Tejas Sapre}, {Mark Shaw}, {Gabriel Silva},
  {Madhan Sivakumar}, {Nisheeth Srivastava}, {Anshuman Verma}, {Qasim Zuhair},
  {Deepak Bansal}, {Doug Burger}, {Kushagra Vaid}, {David A. Maltz}, and
  {Albert Greenberg}.
\newblock Azure {{Accelerated Networking}}: {{SmartNICs}} in the {{Public
  Cloud}}.
\newblock In \emph{15th {{USENIX Symposium}} on {{Networked Systems Design}}
  and {{Implementation}} ({{NSDI}} 18)}, {{NSDI}}'18.
\newblock ISBN 978-1-931971-43-0.

\bibitem[Engler et~al.()Engler, Kaashoek, O’Toole, and Laboratory]{exokernel}
Dawson~R Engler, M~Frans Kaashoek, James O’Toole, and M~I~T Laboratory.
\newblock Exokernel: An {{Operating System Architecture}} for
  {{Application}}-{{Level Resource Management}}.
\newblock page~16.

\bibitem[Fleming()]{ebpf}
Matt Fleming.
\newblock A thorough introduction to {eBPF}.
\newblock URL \url{https://lwn.net/Articles/740157/}.

\bibitem[Frysinger()]{vdso_man}
Mike Frysinger.
\newblock vdso(7).
\newblock URL \url{https://man7.org/linux/man-pages/man7/vdso.7.html}.

\bibitem[Gerhorst et~al.(2021)Gerhorst, Herzog, Reif,
  Schr{\"{o}}der{-}Preikschat, and H{\"{o}}nig]{anycall}
Luis Gerhorst, Benedict Herzog, Stefan Reif, Wolfgang
  Schr{\"{o}}der{-}Preikschat, and Timo H{\"{o}}nig.
\newblock Anycall: Fast and flexible system-call aggregation.
\newblock In \emph{{PLOS} '21: Proceedings of the 11th Workshop on Programming
  Languages and Operating Systems, Virtual Event, Germany, October 25, 2021},
  pages 1--8. {ACM}, 2021.
\newblock \doi{10.1145/3477113.3487267}.
\newblock URL \url{https://doi.org/10.1145/3477113.3487267}.

\bibitem[Ghigoff et~al.()Ghigoff, Sopena, Muller, Lazri, and Blin]{bmc}
Yoann Ghigoff, Julien Sopena, Gilles Muller, Kahina Lazri, and Antoine Blin.
\newblock {{BMC}}: Accelerating {{Memcached}} using {{Safe In}}-kernel
  {{Caching}} and {{Pre}}-stack {{Processing}}.
\newblock page~16.

\bibitem[{Google Inc.}()]{benchmark}
{Google Inc.}
\newblock benchmark.
\newblock URL \url{https://github.com/google/benchmark}.

\bibitem[Gray et~al.(2005)Gray, Chapman, Chubb, Mosberger, and Heiser]{itanium}
Charles Gray, Matthew Chapman, Peter Chubb, David Mosberger, and Gernot Heiser.
\newblock Itanium - {A} system implementor's tale(awarded general track best
  student paper award!).
\newblock In \emph{Proceedings of the 2005 {USENIX} Annual Technical
  Conference, April 10-15, 2005, Anaheim, CA, {USA}}, pages 265--278. {USENIX},
  2005.
\newblock URL
  \url{http://www.usenix.org/events/usenix05/tech/general/gray.html}.

\bibitem[He et~al.()He, Wang, Fu, Tan, Hua, Zhang, and Zheng]{masq}
Zhiqiang He, Dongyang Wang, Binzhang Fu, Kun Tan, Bei Hua, Zhi-Li Zhang, and
  Kai Zheng.
\newblock {{MasQ}}: {{RDMA}} for {{Virtual Private Cloud}}.
\newblock In \emph{Proceedings of the {{Annual}} conference of the {{ACM
  Special Interest Group}} on {{Data Communication}} on the applications,
  technologies, architectures, and protocols for computer communication},
  {{SIGCOMM}} '20, pages 1--14. {Association for Computing Machinery}.
\newblock ISBN 978-1-4503-7955-7.
\newblock \doi{10/gg9rjq}.

\bibitem[Hoefler et~al.()Hoefler, Di~Girolamo, Taranov, Grant, and
  Brightwell]{spin}
Torsten Hoefler, Salvatore Di~Girolamo, Konstantin Taranov, Ryan~E. Grant, and
  Ron Brightwell.
\newblock {{sPIN}}: high-performance streaming processing in the network.
\newblock In \emph{Proceedings of the {{International Conference}} for {{High
  Performance Computing}}, {{Networking}}, {{Storage}} and {{Analysis}} on -
  {{SC}} '17}, pages 1--16. {ACM Press}.
\newblock ISBN 978-1-4503-5114-0.
\newblock \doi{10.1145/3126908.3126970}.

\bibitem[Humphries et~al.()Humphries, Natu, Chaugule, Weisse, Rhoden, Don,
  Rizzo, Rombakh, Turner, and Kozyrakis]{ghost}
Jack~Tigar Humphries, Neel Natu, Ashwin Chaugule, Ofir Weisse, Barret Rhoden,
  Josh Don, Luigi Rizzo, Oleg Rombakh, Paul Turner, and Christos Kozyrakis.
\newblock {{ghOSt}}: Fast \& {{Flexible User}}-{{Space Delegation}} of {{Linux
  Scheduling}}.
\newblock In \emph{Proceedings of the {{ACM SIGOPS}} 28th {{Symposium}} on
  {{Operating Systems Principles}}}, {{SOSP}} '21, pages 588--604. {Association
  for Computing Machinery}.
\newblock ISBN 978-1-4503-8709-5.
\newblock \doi{10/gm8ntm}.

\bibitem[Høiland-Jørgensen et~al.()Høiland-Jørgensen, Brouer, Borkmann,
  Fastabend, Herbert, Ahern, and Miller]{xdp}
Toke Høiland-Jørgensen, Jesper~Dangaard Brouer, Daniel Borkmann, John
  Fastabend, Tom Herbert, David Ahern, and David Miller.
\newblock The {eXpress} data path: fast programmable packet processing in the
  operating system kernel.
\newblock In \emph{Proceedings of the 14th International Conference on emerging
  Networking {EXperiments} and Technologies}, pages 54--66. {ACM}.
\newblock ISBN 978-1-4503-6080-7.
\newblock \doi{10.1145/3281411.3281443}.
\newblock URL \url{https://dl.acm.org/doi/10.1145/3281411.3281443}.

\bibitem[{InfiniBand Trade Association}()]{ibta}
{InfiniBand Trade Association}.
\newblock \emph{{{InfiniBand Architecture Specification}}}, volume~1.
\newblock {InfiniBand Trade Association}, 1.3 edition.
\newblock URL \url{https://cw.infinibandta.org/document/dl/8567}.

\bibitem[{Intel Corporation}({\natexlab{a}})]{fred}
{Intel Corporation}.
\newblock Flexible return and event delivery ({FRED}), {\natexlab{a}}.
\newblock URL
  \url{https://software.intel.com/content/dam/develop/external/us/en/documents-tps/346446-flexible-return-and-event-delivery.pdf}.

\bibitem[{Intel Corporation}({\natexlab{b}})]{hyper-threading}
{Intel Corporation}.
\newblock What is hyper-threading?, {\natexlab{b}}.
\newblock URL
  \url{https://www.intel.com/content/www/us/en/gaming/resources/hyper-threading.html}.

\bibitem[{Intel Corporation}({\natexlab{c}})]{intel_manual}
{Intel Corporation}.
\newblock Intel® 64 and {IA}-32 architectures software developer’s manual,
  combined volumes: 1, 2a, 2b, 2c, 2d, 3a, 3b, 3c, 3d and 4, {\natexlab{c}}.
\newblock URL
  \url{https://software.intel.com/content/dam/develop/public/us/en/documents/325462-sdm-vol-1-2abcd-3abcd.pdf}.

\bibitem[{Intel Corporation}({\natexlab{d}})]{turbo_boost}
{Intel Corporation}.
\newblock What is intel® turbo boost technology?, {\natexlab{d}}.
\newblock URL
  \url{https://www.intel.com/content/www/us/en/gaming/resources/turbo-boost.html}.

\bibitem[Ivanov et~al.()Ivanov, Dryden, Ben-Nun, Li, and
  Hoefler]{ivanovDataMovementAll}
Andrei Ivanov, Nikoli Dryden, Tal Ben-Nun, Shigang Li, and Torsten Hoefler.
\newblock Data {{Movement Is All You Need}}: A {{Case Study}} on {{Optimizing
  Transformers}}.
\newblock page~22.

\bibitem[Jeong et~al.()Jeong, Woo, Jamshed, Jeong, Ihm, Han, and Park]{mtcp}
Eun~Young Jeong, Shinae Woo, Muhammad Jamshed, Haewon Jeong, Sunghwan Ihm,
  Dongsu Han, and {KyoungSoo} Park.
\newblock {mTCP}: a highly scalable user-level {TCP} stack for multicore
  systems.
\newblock In \emph{Proceedings of the 11th {USENIX} Conference on Networked
  Systems Design and Implementation}, {NSDI}'14, pages 489--502. {USENIX}
  Association.
\newblock ISBN 978-1-931971-09-6.

\bibitem[Kanev et~al.()Kanev, Darago, Hazelwood, Ranganathan, Moseley, Wei, and
  Brooks]{kanevProfilingWarehousescaleComputer2015}
Svilen Kanev, Juan~Pablo Darago, Kim Hazelwood, Parthasarathy Ranganathan, Tipp
  Moseley, Gu-Yeon Wei, and David Brooks.
\newblock Profiling a warehouse-scale computer.
\newblock In \emph{Proceedings of the 42nd {{Annual International Symposium}}
  on {{Computer Architecture}}}, {{ISCA}} '15, pages 158--169. {Association for
  Computing Machinery}.
\newblock ISBN 978-1-4503-3402-0.
\newblock \doi{10/ghmjs6}.

\bibitem[Kerrisk and Eckhardt()]{fork}
Michael Kerrisk and Drew Eckhardt.
\newblock fork(2).
\newblock URL \url{https://man7.org/linux/man-pages/man2/fork.2.html}.

\bibitem[Kim et~al.({\natexlab{a}})Kim, Yu, Liu, Zhu, Padhye, Raindel, Guo,
  Sekar, and Seshan]{freeflow}
Daehyeok Kim, Tianlong Yu, Hongqiang~Harry Liu, Yibo Zhu, Jitu Padhye, Shachar
  Raindel, Chuanxiong Guo, Vyas Sekar, and Srinivasan Seshan.
\newblock {{FreeFlow}}: Software-based {{Virtual RDMA Networking}} for
  {{Containerized Clouds}}.
\newblock In \emph{Proceedings of the 16th {{USENIX Conference}} on {{Networked
  Systems Design}} and {{Implementation}}}, {{NSDI}}, pages 113--125,
  {\natexlab{a}}.
\newblock ISBN 978-1-931971-49-2.
\newblock \doi{10.5555/3323234.3323245}.

\bibitem[Kim et~al.({\natexlab{b}})Kim, Lee, and Kim]{nvmedirect}
Hyeong-Jun Kim, Young-Sik Lee, and Jin-Soo Kim.
\newblock {NVMeDirect}: A user-space i/o framework for application-specific
  optimization on {NVMe} {SSDs}.
\newblock {\natexlab{b}}.
\newblock URL
  \url{https://www.usenix.org/conference/hotstorage16/workshop-program/presentation/kim}.

\bibitem[Kocher et~al.()Kocher, Horn, Fogh, Genkin, Gruss, Haas, Hamburg, Lipp,
  Mangard, Prescher, Schwarz, and Yarom]{spectre}
Paul Kocher, Jann Horn, Anders Fogh, \{and\}~Daniel Genkin, Daniel Gruss,
  Werner Haas, Mike Hamburg, Moritz Lipp, Stefan Mangard, Thomas Prescher,
  Michael Schwarz, and Yuval Yarom.
\newblock Spectre attacks: Exploiting speculative execution.
\newblock In \emph{40th {IEEE} Symposium on Security and Privacy (S\&P'19)}.

\bibitem[Kurtzer et~al.()Kurtzer, Sochat, and Bauer]{singularity}
Gregory~M. Kurtzer, Vanessa Sochat, and Michael~W. Bauer.
\newblock Singularity: Scientific containers for mobility of compute.
\newblock 12\penalty0 (5):\penalty0 e0177459.
\newblock ISSN 1932-6203.
\newblock \doi{10/f969fz}.

\bibitem[Lee et~al.(2018)Lee, Song, and Kang]{lotrx86}
Hojoon Lee, Chihyun Song, and Brent~ByungHoon Kang.
\newblock Lord of the x86 rings: {A} portable user mode privilege separation
  architecture on x86.
\newblock In David Lie, Mohammad Mannan, Michael Backes, and XiaoFeng Wang,
  editors, \emph{Proceedings of the 2018 {ACM} {SIGSAC} Conference on Computer
  and Communications Security, {CCS} 2018, Toronto, ON, Canada, October 15-19,
  2018}, pages 1441--1454. {ACM}, 2018.
\newblock \doi{10.1145/3243734.3243748}.
\newblock URL \url{https://doi.org/10.1145/3243734.3243748}.

\bibitem[{LF Projects, LLC}()]{dpdk}
{LF Projects, LLC}.
\newblock Data plane development kit.
\newblock URL \url{https://www.dpdk.org/}.

\bibitem[{Linux Kernel Organization, Inc.}()]{linux}
{Linux Kernel Organization, Inc.}
\newblock The linux kernel archives.
\newblock URL \url{https://www.kernel.org/}.

\bibitem[Lipp et~al.()Lipp, Schwarz, Gruss, Prescher, Haas, Fogh, Horn,
  Mangard, Kocher, Genkin, Yarom, and Hamburg]{meltdown}
Moritz Lipp, Michael Schwarz, Daniel Gruss, Thomas Prescher, Werner Haas,
  Anders Fogh, Jann Horn, Stefan Mangard, Paul Kocher, Daniel Genkin, Yuval
  Yarom, and Mike Hamburg.
\newblock Meltdown: Reading kernel memory from user space.
\newblock In \emph{27th {USENIX} Security Symposium ({USENIX} Security 18)}.

\bibitem[Mi et~al.()Mi, Li, Yang, Wang, and Chen]{skybridge}
Zeyu Mi, Dingji Li, Zihan Yang, Xinran Wang, and Haibo Chen.
\newblock {SkyBridge}: Fast and secure inter-process communication for
  microkernels.
\newblock In \emph{Proceedings of the Fourteenth {EuroSys} Conference 2019},
  {EuroSys} '19, pages 1--15. Association for Computing Machinery.
\newblock ISBN 978-1-4503-6281-8.
\newblock \doi{10.1145/3302424.3303946}.
\newblock URL \url{https://doi.org/10.1145/3302424.3303946}.

\bibitem[{Mike Jackson} and {Ravi
  Budruk}()]{mikejacksonPCIExpressTechnology2012}
{Mike Jackson} and {Ravi Budruk}.
\newblock \emph{{{PCI Express Technology}} 3.0}.

\bibitem[Narayanan et~al.()Narayanan, Huang, Detweiler, Appel, Li, Zellweger,
  and Burtsev]{redleaf}
Vikram Narayanan, Tianjiao Huang, David Detweiler, Dan Appel, Zhaofeng Li, Gerd
  Zellweger, and Anton Burtsev.
\newblock {{RedLeaf}}: Isolation and {{Communication}} in a {{Safe Operating
  System}}.
\newblock pages 21--39.
\newblock ISBN 978-1-939133-19-9.
\newblock URL
  \url{https://www.usenix.org/conference/osdi20/presentation/narayanan-vikram}.

\bibitem[{NVIDIA}()]{nvidiaConnectX6VPICard2020}
{NVIDIA}.
\newblock {{ConnectX}}-6 {{VPI Card Product Brief}}.
\newblock URL
  \url{https://www.mellanox.com/files/doc-2020/pb-connectx-6-vpi-card.pdf}.

\bibitem[{NVM Express, Inc.}()]{nvme}
{NVM Express, Inc.}
\newblock {NVM} express base specification, revision 1.4b.
\newblock URL
  \url{https://nvmexpress.org/wp-content/uploads/NVM-Express-1_4b-2020.09.21-Ratified.pdf}.

\bibitem[Peter et~al.()Peter, Li, Zhang, Ports, Woos, Krishnamurthy, Anderson,
  and Roscoe]{arrakis}
Simon Peter, Jialin Li, Irene Zhang, Dan R.~K. Ports, Doug Woos, Arvind
  Krishnamurthy, Thomas Anderson, and Timothy Roscoe.
\newblock Arrakis: The operating system is the control plane.
\newblock 33\penalty0 (4):\penalty0 11:1--11:30.
\newblock ISSN 0734-2071.
\newblock \doi{10.1145/2812806}.
\newblock URL \url{https://doi.org/10.1145/2812806}.

\bibitem[Planeta et~al.()Planeta, Bierbaum, Antony, Hoefler, and
  Härtig]{migros}
Maksym Planeta, Jan Bierbaum, Leo Sahaya~Daphne Antony, Torsten Hoefler, and
  Hermann Härtig.
\newblock {{MigrOS}}: Transparent {{Operating Systems Live Migration Support}}
  for {{Containerised RDMA}}-applications.
\newblock In \emph{{{USENIX ATC}} 2021}, pages 47--63.
\newblock ISBN 978-1-939133-23-6.
\newblock URL
  \url{https://www.usenix.org/conference/atc21/presentation/planeta}.

\bibitem[Rizzo()]{netmap}
Luigi Rizzo.
\newblock Netmap: a novel framework for fast packet i/o.
\newblock In \emph{Proceedings of the 2012 {USENIX} conference on Annual
  Technical Conference}, {USENIX} {ATC}'12, page~9. {USENIX} Association.

\bibitem[Sadok et~al.()Sadok, Zhao, Choung, Atre, Berger, Hoe, Panda, and
  Sherry]{kopi}
Hugo Sadok, Zhipeng Zhao, Valerie Choung, Nirav Atre, Daniel~S. Berger,
  James~C. Hoe, Aurojit Panda, and Justine Sherry.
\newblock We need kernel interposition over the network dataplane.
\newblock In \emph{Proceedings of the Workshop on Hot Topics in Operating
  Systems}, {HotOS} '21, pages 152--158. Association for Computing Machinery.
\newblock ISBN 978-1-4503-8438-4.
\newblock \doi{10.1145/3458336.3465281}.
\newblock URL \url{https://doi.org/10.1145/3458336.3465281}.

\bibitem[Soares and Stumm()]{flexsc}
Livio Soares and Michael Stumm.
\newblock {FlexSC}: flexible system call scheduling with exception-less system
  calls.
\newblock In \emph{Proceedings of the 9th {USENIX} conference on Operating
  systems design and implementation}, {OSDI}'10, pages 33--46. {USENIX}
  Association.

\bibitem[{Software in the Public Interest, Inc.}()]{bullseye}
{Software in the Public Interest, Inc.}
\newblock Debian "bullseye" release information.
\newblock URL \url{https://www.debian.org/releases/bullseye/index.en.html}.

\bibitem[{The kernel development community}({\natexlab{a}})]{cmd_params}
{The kernel development community}.
\newblock The kernel’s command-line parameters, {\natexlab{a}}.
\newblock URL
  \url{https://www.kernel.org/doc/html/v5.11/admin-guide/kernel-parameters.html}.

\bibitem[{The kernel development community}({\natexlab{b}})]{kpti}
{The kernel development community}.
\newblock Page table isolation ({PTI}), {\natexlab{b}}.
\newblock URL \url{https://www.kernel.org/doc/html/v5.11/x86/pti.html}.

\bibitem[{The kernel development community}({\natexlab{c}})]{linux_doc_mm}
{The kernel development community}.
\newblock Memory management, {\natexlab{c}}.
\newblock URL \url{https://www.kernel.org/doc/html/v5.11/x86/x86_64/mm.html}.

\bibitem[{The kernel development community}({\natexlab{d}})]{mds}
{The kernel development community}.
\newblock Microarchitectural data sampling ({MDS}) mitigation, {\natexlab{d}}.
\newblock URL \url{https://www.kernel.org/doc/html/v5.11/x86/mds.html}.

\bibitem[Torvalds()]{kernel_v5.11}
Linus Torvalds.
\newblock Linux kernel source tree v5.11.
\newblock URL
  \url{https://git.kernel.org/pub/scm/linux/kernel/git/torvalds/linux.git/tree/?h=v5.11}.

\bibitem[Torvalds et~al.()Torvalds, Kleen, and Machek]{entry64}
Linus Torvalds, Andi Kleen, and Pavel Machek.
\newblock 64-bit {SYSCALL} instruction entry.
\newblock URL
  \url{https://elixir.bootlin.com/linux/v5.11/source/arch/x86/entry/entry_64.S}.

\bibitem[{Van Jacobson}()]{vanjacobsonFwdTCP30}
{Van Jacobson}.
\newblock Fwd: {{TCP}} in 30 instructions ({{Was Re}}: Karl {{Auerbach}}: Re:
  Storage over {{Eth}}.
\newblock URL
  \url{https://www.pdl.cmu.edu/mailinglists/ips/mail/msg00133.html}.

\bibitem[Vasudevan et~al.()Vasudevan, Yerraballi, and Chawla]{klos}
Amit Vasudevan, Ramesh Yerraballi, and Ashish Chawla.
\newblock A high performance kernel-less operating system architecture.
\newblock In \emph{Proceedings of the Twenty-eighth Australasian conference on
  Computer Science - Volume 38}, {ACSC} '05, pages 287--296. Australian
  Computer Society, Inc.
\newblock ISBN 978-1-920682-20-0.

\bibitem[Wysocki()]{governor}
Rafael~J. Wysocki.
\newblock {CPU} performance scaling.
\newblock URL
  \url{https://www.kernel.org/doc/html/v5.11/admin-guide/pm/cpufreq.html}.

\bibitem[Yang et~al.()Yang, Harris, Walker, Verkamp, Liu, Chang, Cao, Stern,
  Verma, and Paul]{spdk}
Z.~Yang, J.~R. Harris, B.~Walker, D.~Verkamp, C.~Liu, C.~Chang, G.~Cao,
  J.~Stern, V.~Verma, and L.~E. Paul.
\newblock {SPDK}: A development kit to build high performance storage
  applications.
\newblock In \emph{2017 {IEEE} International Conference on Cloud Computing
  Technology and Science ({CloudCom})}, pages 154--161.
\newblock \doi{10.1109/CloudCom.2017.14}.
\newblock {ISSN}: 2330-2186.

\bibitem[Zhang et~al.(2021)Zhang, Raybuck, Patel, Olynyk, Nelson, Leija,
  Martinez, Liu, Simpson, Jayakar, Penna, Demoulin, Choudhury, and
  Badam]{demikernel}
Irene Zhang, Amanda Raybuck, Pratyush Patel, Kirk Olynyk, Jacob Nelson, Omar
  S.~Navarro Leija, Ashlie Martinez, Jing Liu, Anna~Kornfeld Simpson, Sujay
  Jayakar, Pedro~Henrique Penna, Max Demoulin, Piali Choudhury, and Anirudh
  Badam.
\newblock The demikernel datapath {OS} architecture for microsecond-scale
  datacenter systems.
\newblock In Robbert van Renesse and Nickolai Zeldovich, editors, \emph{{SOSP}
  '21: {ACM} {SIGOPS} 28th Symposium on Operating Systems Principles, Virtual
  Event / Koblenz, Germany, October 26-29, 2021}, pages 195--211. {ACM}, 2021.
\newblock \doi{10.1145/3477132.3483569}.
\newblock URL \url{https://doi.org/10.1145/3477132.3483569}.

\bibitem[Zhong et~al.()Zhong, Wang, Wu, Cidon, Stutsman, Tai, and
  Yang]{bpf_storage}
Yuhong Zhong, Hongyi Wang, Yu~Jian Wu, Asaf Cidon, Ryan Stutsman, Amy Tai, and
  Junfeng Yang.
\newblock {BPF} for storage: an exokernel-inspired approach.
\newblock In \emph{Proceedings of the Workshop on Hot Topics in Operating
  Systems}, {HotOS} '21, pages 128--135. Association for Computing Machinery.
\newblock ISBN 978-1-4503-8438-4.
\newblock \doi{10.1145/3458336.3465290}.
\newblock URL \url{https://doi.org/10.1145/3458336.3465290}.

\end{thebibliography}
\end{document}